\documentclass[a4paper,11pt,final]{article}

\usepackage[colorlinks=true]{hyperref}
\usepackage[english]{babel}
\usepackage[utf8]{inputenc}
\usepackage[T1]{fontenc}
\usepackage{lmodern}
\usepackage[pdftex]{graphicx}
\usepackage[top=2cm, bottom=2cm, left=2cm, right=2cm]{geometry}
\usepackage{setspace}
\usepackage[french]{varioref}
\usepackage{lastpage}
\usepackage{fancyhdr}
\usepackage[table]{xcolor}
\usepackage{tikz}
\usepackage{titling}
\usepackage{graphicx}
\usepackage{amsmath}
\usepackage{amsthm}
\usepackage{amssymb}

\usepackage{multirow}
\usepackage{rotating}
\usepackage{subcaption} 
\usepackage[font=footnotesize,labelfont=bf]{caption}
\usepackage[gen]{eurosym}
\usepackage{dirtree}

\usepackage{rotating}
\usepackage{colortbl}
\usepackage{footmisc}

\usepackage{rotating}
\usepackage{threeparttable, tablefootnote}
\usepackage{multicol}
\usepackage{titlesec}

\usepackage{tabularx}
\usepackage{longtable}
\usepackage{booktabs}
\newcolumntype{C}[1]{>{\centering\let\newline\\\arraybackslash\hspace{0pt}}m{#1}}
\newcolumntype{L}[1]{>{\raggedright\let\newline\\\arraybackslash\hspace{0pt}}m{#1}}
\newcolumntype{R}[1]{>{\raggedleft\let\newline\\\arraybackslash\hspace{0pt}}m{#1}}

\usepackage{pifont} 

\usepackage{doi}
\usepackage[autostyle]{csquotes}
\usepackage[style=numeric, citestyle=numeric-comp, 
	maxcitenames=2, maxbibnames=99, 
    backend=biber
]{biblatex}
\usepackage{graphicx}

\usepackage{enumitem}
\setlist{nolistsep}

\title{FOPPA: an Open Database of French Public Procurement Award Notices From 2010--2020}

\author{Lucas Potin\textsuperscript{1,4}, Vincent Labatut\textsuperscript{1,4}, Pierre-Henri Morand\textsuperscript{2,4} \& Christine Largeron\textsuperscript{3}}

\date{\today \\[0.2cm] 
\textsuperscript{1}: Laboratoire Informatique d'Avignon UPR 4128, Avignon Université, Avignon, France \\
\textsuperscript{2}: Laboratoire Biens Normes et Contrats UPR 3788, Avignon Université, Avignon, France \\
\textsuperscript{3}: Laboratoire Hubert Curien UMR 5516, Université Jean Monnet, Saint-\'Etienne, France \\
\textsuperscript{4}: Agorantic, FR 3621, Avignon Université, Avignon, France
}

\hypersetup{
    pdftitle={\thetitle},
    pdfauthor={\theauthor},
    bookmarksnumbered=true,bookmarksopen=true,
	unicode=true,colorlinks=true,linktoc=all,
	linkcolor=blue,citecolor=blue,filecolor=blue,urlcolor=blue,
	pdfstartview=FitH
}

\begin{document}
\maketitle

\begin{abstract} 
\textit{Public Procurement} refers to governments' purchasing activities of goods, services, and construction of public works. In the European Union (EU), it is an essential sector, corresponding to $15 \%$ of the GDP. EU public procurement generates large amounts of data, because award notices related to contracts exceeding a predefined threshold must be published on the TED (EU's official journal). 
Under the framework of the DeCoMaP project, which aims at leveraging such data in order to predict fraud in public procurement, we constitute the FOPPA (French Open Public Procurement Award notices) database. It contains the description of $1{,}380{,}965$ lots obtained from the TED, covering the 2010--2020 period for France. We detect a number of substantial issues in these data, and propose a set of automated and semi-automated methods to solve them and produce a usable database. 
It can be leveraged to study public procurement in an academic setting, but also to facilitate the monitoring of public policies, and to improve the quality of the data offered to buyers and suppliers.


\medskip
\noindent\textcolor{red}{\textbf{Cite as:} L. Potin, V. Labatut, P.-H. Morand \& C. Largeron. FOPPA: an Open Database of French Public Procurement Award Notices From 2010--2020, in \textit{Scientific Data}, 2023, 10:303. DOI: \href{https://www.doi.org/10.1038/s41597-023-02213-z}{\texttt{10.1038/s41597-023-02213-z}}}

\end{abstract}

\section{Background \& Summary} 

Public Procurement refers to governments' purchasing activities of goods, services, and construction of public works. It comprises large shares of government budgets and gross domestic products (GDP) in the majority of the countries in the world. For instance, such contracts are believed to amount to some \euro{}200 billion a year in France, corresponding to $10 \%$ of the country's GDP~\cite{Saussier2015}.

As one of the most important areas where the state and the private sector interact extensively, public procurement procedures are open to the use of public resources for different interests other than the public good. Their implementation may well involve corrupt transfers between government officials and private sector firms, for reasons ranging from personal interests to the financing of political parties~\cite{Guerakar2016}. The World Bank has estimated that roughly \$1.5 trillion in public contracts are influenced by corruption~\cite{Passas2007}, and that the volume of bribes exchanging hands for public procurement alone is about \$200 billion per year~\cite{Kaufmann2005}. The World Business Environment Survey carried out by the World Bank in 2000 shows that about $60 \%$ of the companies admitted giving bribes~\cite{Boehm2006}.

According to the Organization for Economic Co-operation and Development (OECD)~\cite{Ubaldi2017}, open data can help to design better anti-corruption policies and to monitor their effective implementation. First, they constitute an incentive to avoid illegal behavior by increasing the chance of exposing governmental misconduct. Second, they ease the design of tools and mechanisms that can be used by judicial institutions to discover and dismantle corrupt activities. Third, they can be leveraged by media and society to detect abuse. 
This focus on open data requires the development of regulations, hence the new public procurement rules adopted in 2014\footnote{Pursuant to European Directive 2014/24/EU on public procurement, Directive 2014/25/EU on procurement by entities operating in the water, energy, transport and postal services sectors, and Directive 2014/23/EU on the award of concession contracts} explicitly state that contracting authorities should store and provide access to data describing high-value contracts as well as their awarding procedure, in an online, interoperable and convenient way. It was in that vein that France announced her willingness to support the implementation of the Open Contracting Data Standard during the London G20 Anti-Corruption Summit in May 2016~\cite{Savy2017}. This standard aims to foster public sector transparency, and to fight corruption and nepotism on public procurement processes by following an open by-default approach during the whole public contracting process.

The DeCoMaP project (\textit{Détection de corruption dans les marchés publics} -- Detection of corruption in public procurement markets) lies at the intersection of these three considerations: regulatory developments, open data, and automatic tools for fraud detection and economics analysis. It involves an interdisciplinary team that includes academic researchers in Economics, Law, and Computer Science, as well as operators of the open data industry. The project, which is funded by the ANR (French national research agency) under grant number ANR-19-CE38-0004, started in 2019 and will end in 2024. It mainly has three goals: 1) constitute a database of public procurement notices and perform its descriptive analysis from the perspective of both Economics and Law; 2) leverage this database to propose an automatic and explainable method able of detecting fraud in public procurement; 3) take advantage of the descriptive analysis and the explainability of the detection method to provide the public government with guidelines regarding the improvement of public procurement open data. 


At the first stage of the DeCoMaP project, we constituted a database of French public procurement notices, whose description is the object of the present article. It is called FOPPA, for \textit{French Open Public Procurement Award notices}. During the elaboration of FOPPA, we detected a number of serious issues in the raw open data provided by the European Union: missing values (in particular unique identifiers), incorrect values, mixed fields. Although common, such issues regarding the quality of open data are practically never mentioned in the public procurement literature, and authors rarely explain how they solve them. This article is the occasion to report the nature and extent of these problems, and the methods that can be applied to solve them, by considering the work we conducted to constitute FOPPA.

\section{Methods}
The raw data describing French public procurement is publicly available online through a European Union (EU) service called TED, which we describe later. However, they exhibit a number of issues that must be solved before performing any analysis. Identifying and correcting these issues is far from being a trivial task. In their study of the whole TED dataset, Csáki \& Prier~\cite{Csaki2018} even stress that ``given its size and complexity, this dataset is not for the faint of heart and [...] no current quality frameworks appear readily able to help''. In this section, we first describe the characteristics of the raw data, and summarize their main issues. Then, we present the processing that we propose to fix these issues.

\subsection{Raw Data and Their Issues}
We first introduce a few notions related to public procurement, and required to properly understand the nature of the data and of our proposed correction process. We then focus on the raw data themselves, before listing their main issues.

\subsubsection{Public Procurement and Related Notions}
Public procurement refers to the purchase of goods, services and works by a public authority (the \textit{buyer}) from a legal entity governed by public or private law (the \textit{supplier}). We collectively refer to buyers and suppliers as \textit{agents}. In this work, we focus on the case of French public procurement. Public procurement must follow a specific set of rules defined by law, and aiming at preserving freedom of access for all  potential candidates, the equal treatment of their offers, and the transparency of the awarding procedure.

The most common procedure goes as follows. First, the public authority identifies its needs, and separates them into several parts called \textit{lots}. If the estimated value of the lot is above the so-called \textit{European Threshold}~\cite{Grow2021}, then the public authority must publish a call for tenders on the TED, that ends after a so-called \textit{acceptance period}. The exact value of European threshold depends on several factors, in particular the nature and activity domain of the concerned public authority (state, local government, health institution...), and the nature of the considered contract (goods, services, works). Moreover, these thresholds are revised every two years by the European Commission. For the 2020--2021 period, they range from 139 k€ to 5.35 M€ (see our technical report~\cite{Potin2022} for more detail). Among other things, the call for tenders specifies the \textit{criteria} used to select the winner of the call. They must include the contract value, and can add criteria related to the object of the contract or its implementation (e.g. technical value, innovative nature, staff experience, delivery times...). At the end of the acceptance period, the public authority studies the received offers and decides whether to award the different lots to one or more candidate suppliers, who are called \textit{winners}. The public authority indicates its choice with a \textit{contract award notice}, which must also be published on the TED.

\subsubsection{TED Dataset}
The \textit{Tenders Electronic Daily} (TED) is the online version of the EU official journal that is dedicated to the publication of the contract notices (i.e. calls for tenders) and award notices related to public procurement. These data are accessible on data.europa.eu, the official portal for European data, under two possible forms: directly through an API~\cite{Ted2019b}, and as aggregated CSV files~\cite{Ted2022}. They cover a period ranging from 2006 to the present day. However, the notices published in the TED between 2006 and 2009 are both less complete and less reliable~\cite{TED2021}. The latter date also corresponds to a change in the database structure, that brought a number of new fields, among which crucial information such as the national identifier of the winner and the weight associated with the price criterion. Finally, the typology of the activity domains mentioned in these notices was revised in 2009, too. For these three reasons, we focus on the 2010--2020 period~\cite{Ted2010, Ted2010a, Ted2011, Ted2011a, Ted2012, Ted2012a, Ted2013, Ted2013a, Ted2014, Ted2014a, Ted2015, Ted2015a, Ted2016, Ted2016a, Ted2017, Ted2017a, Ted2018, Ted2018a, Ted2019, Ted2019a, Ted2020, Ted2020a}.

For this period, the TED contains $2{,}106{,}606$ award notices, corresponding to $7{,}169{,}070$ lots. In the context of DeCoMaP though, we focus only on the French contracts, amounting to $410{,}283$ award notices ($19.5 \%$), and $1{,}380{,}965$ lots ($19.2 \%$). France is the second member state in terms of number of lots, right behind Poland, and largely ahead of Romania (more than three times the number of lots).

Each row in the TED table represents a specific lot, which is described through $75$ distinct fields. We distinguish four categories of fields: notice metadata (e.g. date of publication, current status, number of corrections); agent information (e.g. national identifier, name and address of the buyer and winner); lot information (e.g. contract type, activity domain, criteria and their relative weights) and award information (e.g. date of the contract award, number of offers, value of the contract as eventually agreed).

\subsubsection{Main Issues}
When assessing the quality of the raw TED data, we identified five issues that we summarize here (see our technical report for a more detailed description, with examples~\cite{Potin2022}). The first one concerns \textit{missing notices}. For a given contract, the TED is likely to contain both a contract notice and an award notice. Most of the time, both should be present, although there are a number of exceptions where either the contract or award notice is optional. Out of the $478{,}854$ contract notices, $186{,}607$ ($39 \%$) do not have a matching award notice, without it being unambiguously optional. According to the practices observed on the TED~\cite{TED2020b}, most of these cases are assumed to be unsuccessful procedures not declared explicitly, for which the award notice is optional. Unfortunately, they are indistinguishable from errors (i.e. truly missing award notices). Out of the $410{,}283$ award notices, $76{,}279$ ($19 \%$) are not cited in any contract notice, without it being explicitly optional. Like before, this number includes actual errors, i.e. where the contract notice should be there, but also cases where it is optional, e.g. renewal of a framework agreement. For both types of notices, these values are in line with the findings of another study conducted on the whole TED dataset for a different period~\cite{Csaki2018}. It is worth stressing that award notices cover most of the fields present in the contract notices, and only a few important fields such as the contract duration and advertisement period are lost when their matching contract notices are missing.

The second issue is \textit{joint agent description}. In the TED table, each row represents a single lot, including dedicated fields to store the information describing the buyer (name, street, zipcode...), and similarly for the winner. However, in practice, it is possible for a lot to be awarded by \textit{several} buyers to \textit{several} winners. This concerns $34{,}257$ ($2 \%$) and $37{,}443$ lots ($3 \%$), respectively. This problem was also mentioned in a previous study dealing with the whole TED dataset on a different period, although it did not quantify its extent~\cite{Csaki2018a}. In this case, all the concerned buyers are described in the same field, under the form of a single character string, and the same holds for the winners. Some data entry clerks appear to use separators characters such as \texttt{-\/-\/-}, but this is neither systematical nor consistent. In order to take advantage of these data, we need a separate representation of these agents.

The third issue is \textit{name and address inconsistency}. The same agent can appear under different names in the dataset, as these strings are not normalized: diacritics and punctuation signs are not used consistently, acronyms are used non-systematically in place (or in combination) of full names. Sometimes, the name field contains additional information related to the physical location of the agent (ex. building number), or its role in a larger structure (ex. internal department). The same type of problem occurs for addresses: lack of normalization, and field pollution (e.g. PO box number appearing in the city field). In addition, the data mix three types of addresses: geographic, postal, and geopostal. All of this makes it impossible to directly perform exact matching between these strings, both within the TED or with external sources.

The fourth issue is \textit{unconstrained criterion description}. The TED lists the award criteria and their respective weights, but the way this information is structured makes it difficult to use, as clerks do not always adopt the same convention to fill the concerned fields. First, as for agent descriptions, multiple criterion names are sometimes inserted together in the same field ($3 \%$ of the lots), and sometimes with inconsistent strings used as separators. Second, the weight of the price criterion, which is supposedly indicated in a separate field, is sometimes mixed with the other criteria ($58 \%$). Third, the weights are not normalized: their bounds are not fixed, and they can sum to any value ($13 \%$). Fourth, sometimes the criteria and their weights are filled together in the same field ($2 \%$). All these problems are important to us, because award criteria are likely to constitute a discriminant information in the context of corruption or fraud prediction~\cite{Fazekas2016}.

The fifth issue is the most serious by far, and concerns \textit{missing agent identification}. The TED contains a unique identifier to refer unambiguously to each economic agent. In the case of France, it is a 14-digit number called SIRET (\textit{Système Informatique pour le Répertoire des Entreprises sur le Territoire} -- Computer system for the national register of companies). It is filled in only $16.4 \%$ of the lots for the buyers, and $2.9 \%$ for winners. An examination of the whole TED dataset reveals that the same observation applies to the rest of the data, with $32.5 \%$ and $12.9 \%$ filled SIRETs for buyers and winners, respectively. On this point, France is among the worst 5 countries, overall. In DeCoMaP, we want to extract various types of procurement networks, considering buyers and winners as vertices. Therefore, correctly identifying all instances of the same agent is absolutely crucial. To solve this problem, we need to leverage all the agent-related information available in the TED: name, location, and activity domain. However, if the corresponding fields are filled most of the time for buyers, they are missing for winners in $25 \%$ of the lots. This difference could be due to the fact that buyers are the one completing the notices, and therefore better fill their parts. The activity domain is more difficult to handle, because it is not described at the level of agent, but rather at the level of the lot. As we will see later, solving this problem requires leveraging an external data source.

\subsection{Data Processing}
\begin{figure}[ht]
    \centering
    \includegraphics[width=\linewidth]{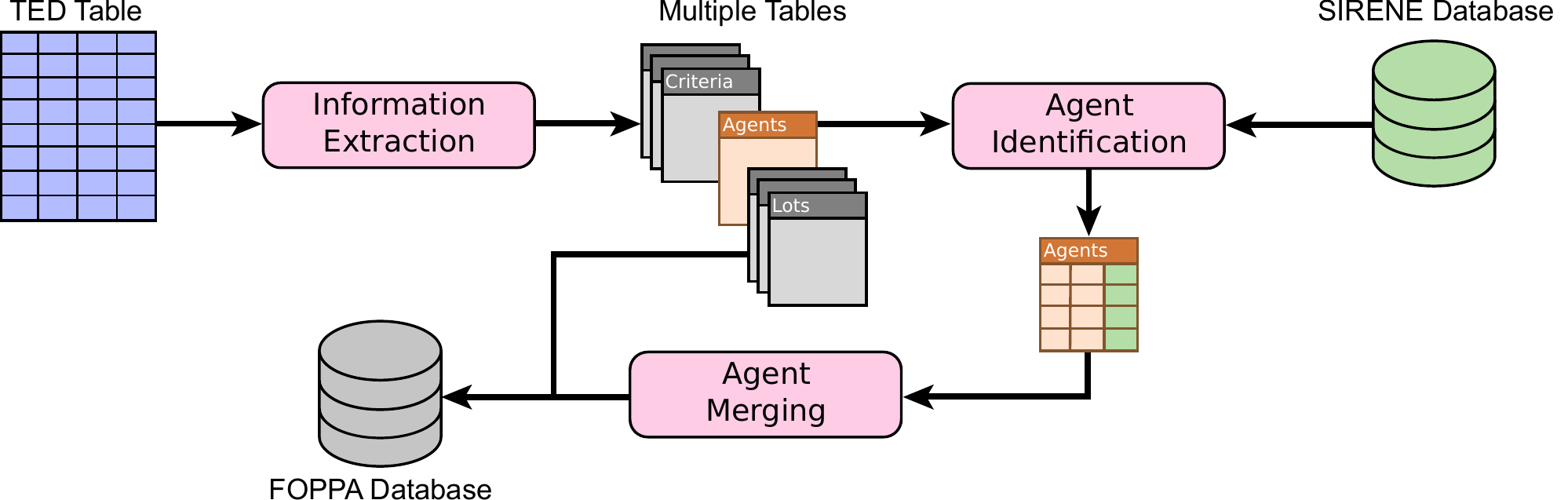}
    \caption{Overview of the method proposed to correct and complete the raw TED data and constitute the FOPPA database.}
    \label{fig:OverallProcess}
\end{figure}

As illustrated in Figure~\ref{fig:OverallProcess}, our process includes three main steps. The first one consists in performing information extraction in order to split certain TED fields that contain heterogeneous information. The second aims at recovering the unique identifier that is missing for most agents described in the database. Finally, the third step concerns the matching and merging of distinct agent occurrences that correspond to the same unique agent. This section summarizes these steps, and their performance is assessed later in Section \textit{Technical Validation}. As indicated before, the interested reader will find more details in our technical report~\cite{Potin2022}. 

\subsubsection{Database Initialization}
The goal of this first step is to split the single original TED CSV table into several separate database tables (cf. the \textit{Data Records} section for a summary of these tables), in order to ease both the cleaning and usage of the data. This requires a series of operations, which are summarized below.

We perform five operations to solve the unconstrained criterion description issue. First, we clean the weight fields by keeping only numerical values and separator strings that we manually identify beforehand. Second, we leverage these separators to split multiple criteria, both regarding their names and their weights. Third, when criterion names and weights are mixed in the same text field, we separate them. Fourth, we normalize the weights of each lot so that they sum to $100$. Fifth, in order to handle the diversity of criterion names due to this field being free text, we add a normalized version of these names, taking the form of one of six possible predefined classes (price, deadline, technical terms, environmental terms, social terms, others).

The rest of the operations concerns the agents. We deal with the address inconsistency issue by removing all irrelevant information from the zipcode and city fields. In addition, we leverage Hexaposte, the French online postal service database~\cite{LaPoste2022}, to fill missing zipcodes. We fix typographical inconsistencies in all address fields by removing all punctuation marks, collapsing consecutive space characters, and switching the remaining text to upper case. In order to deal with the confusion between postal and geographic addresses, we remove words related to the former (e.g. PO numbers).

We proceed similarly to normalize agent names, i.e. punctuation removal, consecutive spaces collapse, and upper case switching. In addition, we remove the text in parentheses, which appears sporadically and corresponds to superfluous indications. We also deal with the joint agent description issue here, by leveraging previously identified separator strings to split the concerned fields (name, address, city, etc.). After splitting all concerned buyer and winner names, we obtain $255,128$ additional agent occurrences ($+9 \%$). Finally, we merge instances that differ in the way they are described (e.g. dissimilar names) but actually correspond to the same entity according to their SIRET (French unique identifier).

\subsubsection{Agent Identification}
The goal of this second step is to solve the missing agent identification issue, which is a major limitation of the data, as explained before. For this purpose, we leverage the SIRENE database~\cite{Sirene2021} (\textit{Système National d'Identification et du Répertoire des Entreprises et de leurs Etablissements} -- National identification system for commercial entities and their facilities), which lists all economic agents participating in public procurement in France. It includes not only agents that are currently active, but also agents that are no longer active, since 1997. This database actually distinguishes two levels of economic agents: entities vs. facilities. \textit{Entities} are companies, government agencies, departments, charities, institutions (legal entities) or people (natural persons) that have a legal existence and the ability to enter into agreements or contracts. \textit{Facilities} are geographically located units where all or part of the entity's economic activity is carried out. Each entity is identified through a unique 9-digit number called the SIREN (\textit{Système d'Identification du Répertoire des Entreprises} -- Identification system of the entity register), whereas for a facility it is a 14-digit number called the SIRET (\textit{Système d'Identification du Répertoire des Etablissements} -- Identification system of the facility register). The first 9 digits of the SIRET correspond to the SIREN of the associated entity, while the last 5 digits are called the NIC (\textit{Numéro Interne de Classement} -- Internal classification number) and are specific to each facility. Two facilities linked to the same entity share the same SIREN, but have a different NIC, and therefore a different SIRET. Agents from the TED correspond to facilities: ideally, we want to identify their SIRETs, but getting their SIRENs right is much better than nothing. 

\begin{figure}[ht]
    \centering
    \includegraphics[width=\textwidth]{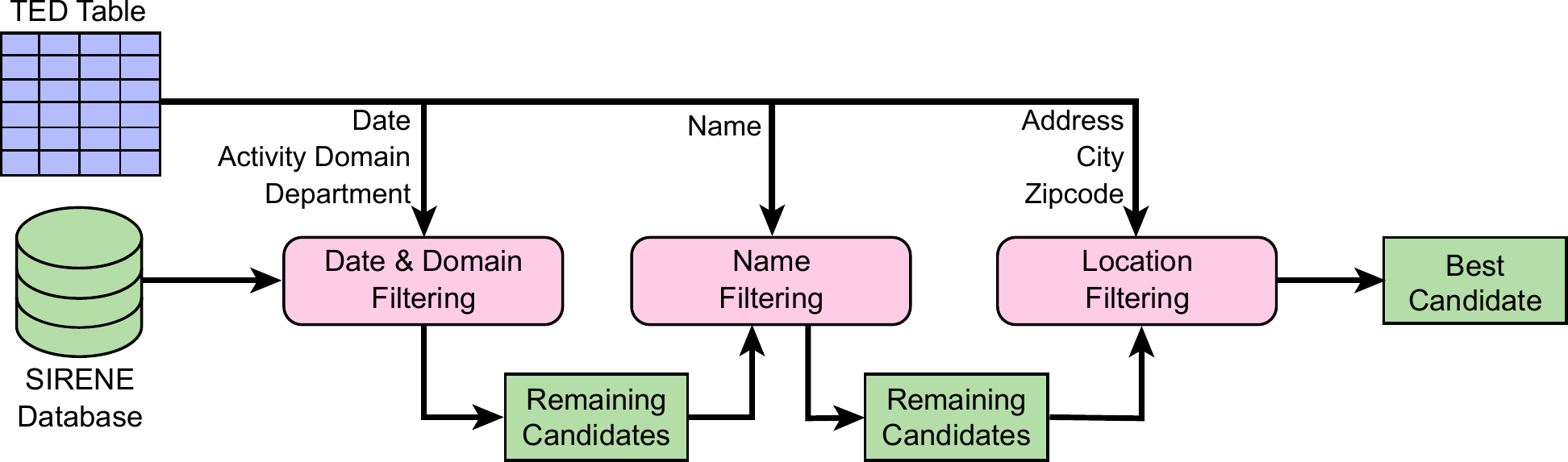}
    \caption{Successive filtering phases applied to match TED agents with SIRENE entries, as a part of the \textit{Agent Identification} step from Figure~\ref{fig:OverallProcess}.}
    \label{fig:Siretization}
\end{figure}

To retrieve the SIRETs from SIRENE, we must match each concerned TED agent to a SIRENE entry. One could assume that using the sole agent name would be the best approach for this purpose. However, this is not that simple, as names can be quite different in both databases (e.g. full name in one vs. acronym in the other). Moreover, SIRENE typically proposes several alternative and/or former names for one agent. For this reason, we propose a filter-based approach illustrated in Figure~\ref{fig:Siretization}, that leverages all the information available in both datasets, i.e. the name and full address of the agents, and the date and activity domain of the lots. It is designed to handle the fact that, as explained before, these fields themselves are not always filled. When searching SIRENE, we consider the information at the level of the facilities, but also, secondarily, of their parent entities, in order to increase the chance of matching. 

We first constitute a set of candidates by selecting the SIRENE entries that are consistent with the date and activity domain of the considered TED lot, as well as the department (a French administrative subdivision) of the TED agent. This allows greatly reducing the number of comparisons in the rest of the process. Second, we filter these candidates by performing an approximate comparison of their names. Third, we filter further by considering the location of the candidates. Based on a scheme that takes into account various types of approximate string matching and the presence/absence of certain parts of the address, our method computes a relevance score for each remaining candidates. In the end, we select the highest ranking candidate as the match. At each step, it is possible that no candidate at all remains, in which case our method returns no match, and the agent identification process therefore fails.

\subsubsection{Agent Merging}
Our agent identification method is not able to retrieve all missing agent SIRETs, which means that a part of the TED dataset remains unidentified after the previous step. We assume that some of these occurrences represent the same agents as other successfully identified occurrences, or even other unidentified occurrences. The goal of this step is to group such occurrences under the same entries in our database, based on their similarity.

We leverage the Dedupe library~\cite{Gregg2022} in order to obtain a clustering of the agent occurrences, comparing them based on their name, address, city and zipcode. We then need to handle the resulting clusters of similar occurrences, of which we identified four types. First, some of the resulting clusters are singletons, meaning that we did not find any other similar occurrence. If the agent in question has no identifier, then our method fails to infer its SIRET and we use an internal code instead. Second, some clusters contain occurrences possessing different SIRETs, in which case we have a conflict. We determined empirically that assigning the majority SIRET to all the occurrences in the cluster is the most effective strategy. Third, some clusters contain only occurrences with no SIRET: we consider that they all correspond to the same agent, and assign the same internal code to all of them. Fourth and finally, certain clusters contain one identified occurrence and one or more agent occurrences with no SIRET: we assign the only known SIRET to all the other occurrences.

After processing the clusters, we have a number of agent occurrences with the same SIRET but not exactly the same information. We merge these entries of our database in order to keep as much of the available information as possible. In case of conflicting fields, we use the same majority rule as above.

\section{Data Records}
\label{sec:DataRecords}
The FOPPA database is publicly available online as a Zenodo\footnote{\url{https://doi.org/10.5281/zenodo.7808664}} repository~\cite{Potin2022a}. It contains an SQL dump of the full database, as well as a CSV version of each table constituting this database.

\begin{figure}[ht]
    \centering
    \includegraphics[width=0.9\textwidth]{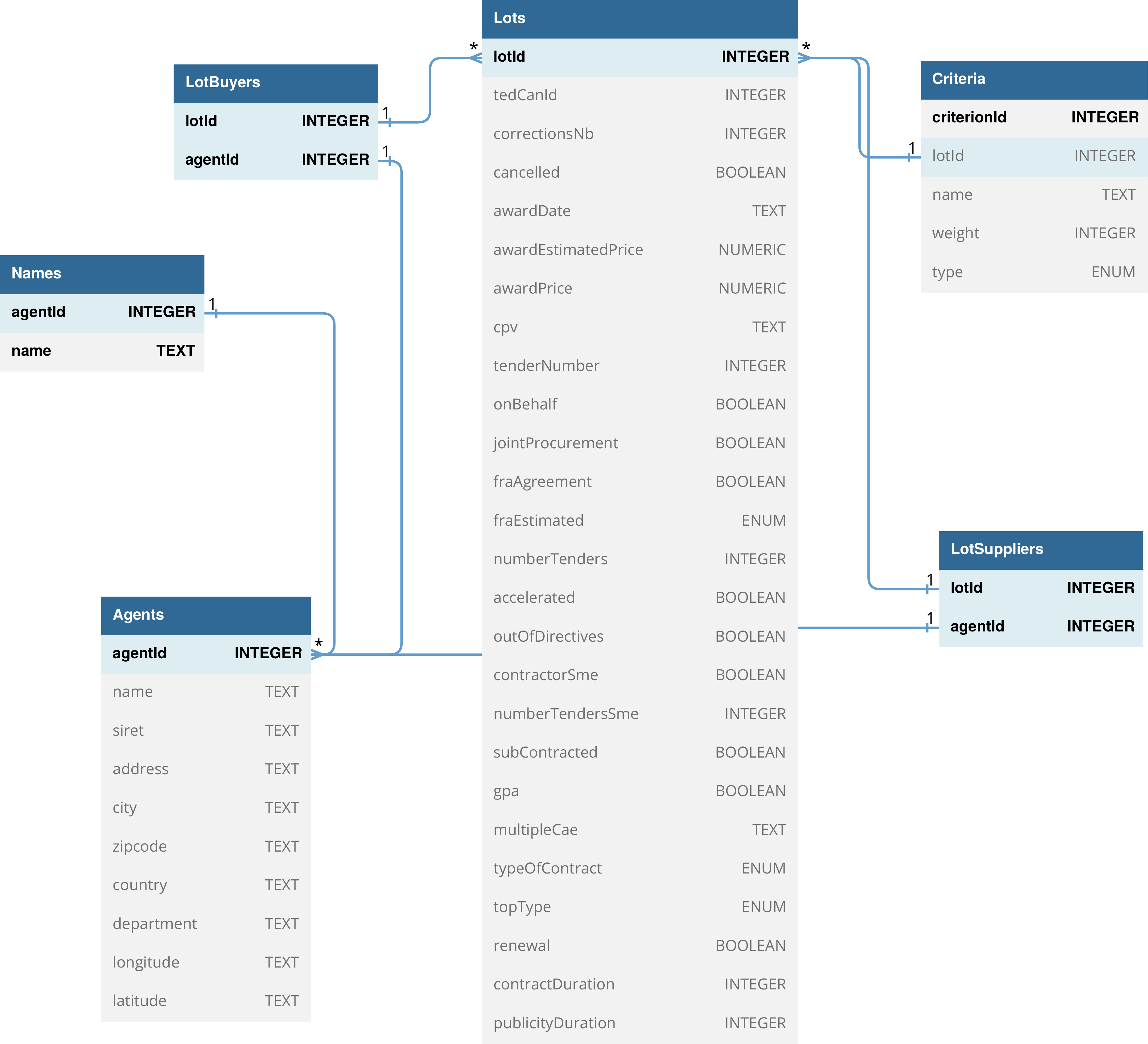} 
    \caption{Structure of the FOPPA database, shown as an Entity-Relation diagram.}
    \label{fig:Database}
\end{figure}

Figure~\ref{fig:Database} shows the dependencies between these tables under the form of an entity-relation diagram. They contain the following information:
\begin{itemize}
    \item \texttt{Lots}: the main table, each entry describes a lot from the TED dataset.
    \item \texttt{Agents}: contains all the economic agents involved in these lots, as buyers (public authorities) or winners (suppliers).
    \item \texttt{Names}: one agent can have several names, which are listed in this table.
    \item \texttt{LotBuyers} \& \texttt{LotSuppliers}: describes the involvement of agents in lots, respectively as buyers and suppliers. One agent can be simultaneously a buyer for a lot and a supplier for a different lot.
    \item \texttt{Criteria}: each lot can have several award criteria, which are listed in this table, together with their associated weight.
\end{itemize}

After the processing described in the \textit{Methods} section, the FOPPA database contains $1{,}380{,}965$ lots, corresponding to $410{,}283$ contract award notices, and involving $301{,}096$ unique agents. Table~\ref{tab:Summary} summarizes the main properties of the database.

\begin{table}[ht]
    \centering
    \begin{tabularx}{\textwidth}{l X}
        \hline
        \textbf{Subject} & Public procurement \\ 
        \textbf{Stored information} & Contract award notices \\
         & Economical agents (buyers \& winners) \\ 
        \textbf{Geographical scope} & France \\ 
        \textbf{Temporal scope} & 2010--2020 \\ 
        \textbf{Size} & $1{,}380{,}965$ lots \\
         & $410{,}283$ contract award notices \\
         & $301{,}096$ agents \\
        \textbf{Data source} & TED (Tenders Electronic Daily) \\ 
        \textbf{Potential uses} & Fraud detection, red flag computing, connection to a number of French databases \\
        \textbf{URL} & \url{https://doi.org/10.5281/zenodo.7808664} \\ 
        \hline
    \end{tabularx}
    \caption{Main properties of the FOPPA database.}
    \label{tab:Summary}
\end{table}

\section{Technical Validation}
As explained in the \textit{Methods} section, there are a number of issues with the raw TED data describing the French public procurement notices. We designed and applied several methods in order to solve most of these issues. In this section, we assess their performance. The operations described in the \textit{Database Initialization} section are straightforward, so we focus on \textit{Agent Identification} (retrieving the missing agent SIRETs) and \textit{Agent Merging} (merging agent occurrences that correspond to the same entity).

\subsection{Identification Method}
In this section, we focus on our method designed to retrieve the missing agent SIRETs. We assess its performance by comparing its output with two distinct ground truths. On the one hand, we leverage TED entries whose SIRET is already filled out in the TED. However, the agents concerned by these cases are mainly buyers, which suggests a potential bias. This is why, on the other hand, we consider a random sample of entries originally without SIRET, which we identify manually.

In order to assess the performance of our method, we consider four different outcomes when matching the agents from our database to the entries of the SIRENE database:
\begin{enumerate}
    \item \textit{Full SIRET}: the matching between our agent and SIRENE is successful, and the method therefore correctly retrieves the whole SIRET, i.e. all 14 digits.
    \item \textit{Partial SIRET}: the method manages to match our agent to the appropriate SIRENE \textit{entity}, but fails to identify the exact \textit{facility}. Therefore, only the SIREN part of the retrieved SIRET is correct, i.e. the first 9 digits.
    \item \textit{Incorrect SIRET}: the method matches our agent to an incorrect entry of SIRENE. Therefore, it returns a completely incorrect SIRET, in the sense that this identifier refers to a totally different agent.
    \item \textit{No SIRET}: the method fails to match our agent to any SIRENE entry, and thus returns no SIRET at all.
\end{enumerate}

\begin{figure}[ht]
    \centering
    \includegraphics[width=0.8\textwidth]{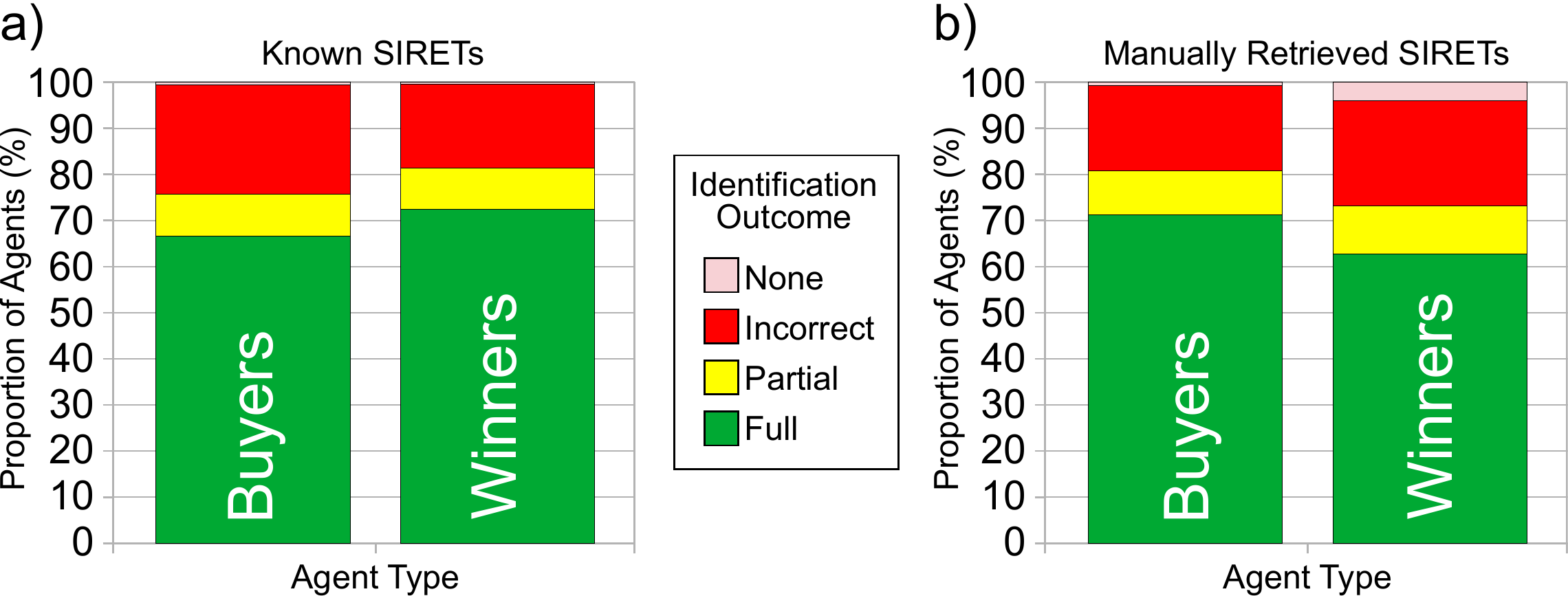}
    \caption{Distribution of the four possible outcomes of the identification process for the \textit{known} (a) and \textit{manually retrieved} (b) SIRETs.}
    \label{fig:ResSiretsTED}
\end{figure}

Figure~\ref{fig:ResSiretsTED} shows the results obtained for both ground truths: known SIRETs (\ref{fig:ResSiretsTED}a) vs. manually retrieved SIRETs (\ref{fig:ResSiretsTED}b). The $x$-axis represents the agents type: the left-hand bar focuses on the buyers and the right-hand one on the winners. Each color corresponds to an outcome: \textit{Full SIRET} (green), \textit{Partial SIRET} (yellow), \textit{Incorrect SIRET} (red), and \textit{No SIRET} (pink). The $y$-axis represents the percentage of agent occurrences for each outcome.

\subsubsection{Known SIRETs}
If we count the number of agents with a valid SIRET, we get $6{,}115$ unique buyers (by opposition to occurrences) and $18{,}209$ unique winners. We remove these SIRETs from the database and apply our method, to check whether it can recover them. 

The complete identification of the SIRET can be performed for $66.6 \%$ of the unique buyers and $72.4 \%$ of the unique winners. For buyers, we consider that a partial SIRET is also a good result, since procurement management is often centralized at the main entity. Summing up the two, we get $75.7 \%$ of success for buyers and $81.4 \%$ for winners. 
For this ground truth, the performance is a bit lower for the buyers, probably because they are the ones filling the award notices. When they bother filling their own information, they tend to get their SIRET right, but not pay too much attention to the rest of their fields. On the contrary, when they fill the winner's SIRET, they generally tend to make sure that they get its other fields right, too.

\subsubsection{Manually Retrieved SIRETs}
To constitute the second ground truth, we first randomly sample $500$ distinct agents ($250$ buyers and $250$ winners) from the TED entries whose SIRET is missing, but that possess both a city and a name. We then take advantage of these two fields to manually retrieve from SIRENE the missing SIRETs of all $500$ agents. This allows solving the potential bias identified in the previous ground truth.

For full SIRETs, on unique buyers we get a slightly better performance than previously (i.e. with the known SIRETs ground truth), with $71.2 \%$. On the contrary, it is a bit lower for winners, with $62.8 \%$ correctly identified SIRETs. When summing the full and partial SIRETs scores, we reach levels comparable to the previous results: $80.8 \%$ for buyers and $73.2 \%$ for winners. This shows that our method generalizes well to new data.

The proportion of agents for which the algorithm is not able to return any SIRET is a bit larger than with the first ground truth, though. This is because these agents originally have many missing fields in the TED and/or a poorly written name (compared to the one present in SIRENE). 


\subsection{Merging Method}
In this section, we assess the performance of our merging step, which aims at combining agents that are dissimilar but correspond to the same entity. This concerns not only agents possessing a SIRET, but also those without any. The clustering process distributes the $306{,}984$ remaining agents over $301{,}096$ clusters. These are small, with an average size of $1.08$ agents by cluster. Table~\ref{tab:clusterDis} shows the full distribution of the cluster sizes, and it appears that most of them are singletons ($94 \%$).

\begin{table}[ht]
    \centering
    \begin{tabularx}{\textwidth}{X r r r r r r r r r r r}
        \hline
        \textbf{Cluster size} & \textbf{1} & \textbf{2} & \textbf{3} & \textbf{4} & \textbf{5} & \textbf{6+} & \textbf{Total} \\
        \hline
        \textbf{Count} & $296{,}296$ & $4{,}158$ & $438$ & $118$ & $41$ & $45$ & $301{,}096$ \\ 
        \textbf{Proportion} & $98.40\%$ & $1,38\%$ & $0.14\%$ & $0.06\%$ & $0.01\%$ & $0.01\%$ &  $100\%$ \\ 
        \hline
    \end{tabularx}
    \caption{Distribution of the number of agents by cluster, for the whole dataset.}
    \label{tab:clusterDis}
\end{table}

As explained in the \textit{Agent Merging} section (first case discussed), singleton clusters do not require any additional processing during the post-processing: if they have a SIRET, then it is assumed correct, and if they do not, it means that we could not find one. The remaining $4{,}800$ clusters contain several agents, possibly corresponding to a single or several identifiers (be them SIRETs or SIRENs), or none at all. There are $3{,}084$ clusters containing conflicting identifiers, which correspond to the second case discussed in the \textit{Agent Merging} section. There are $1{,}099$ clusters containing several unidentified agents (third case), and $617$ clusters containing several agents including only a single identified one (fourth case).

We call \textit{false positives} agents placed in the same cluster, when they actually correspond to several distinct entities, and should therefore be separated. We call \textit{false negatives} agents placed in different clusters, when they actually correspond to the same entity, and should therefore be gathered. In the following, we discuss the performance of our method separately for both types of error.

\subsubsection{False Positives}
False positives correspond to cases of conflicting SIRETs in the same cluster. Table~\ref{tab:clusterSiretsDiff} represents the distribution of clusters according to their numbers of distinct SIRETs. By comparison, Table~\ref{tab:clusterDis} focuses on agents, not SIRETs (some agents have no SIRET). It appears that $298{,}012$ ($98 \%$) clusters contain 0 or 1 SIRET, which means they exhibit no SIRET conflict.

\begin{table}[ht]
    \centering
    \begin{tabularx}{\textwidth}{X r r r r r r r}
        \hline
        \textbf{Number of distinct SIRETs} & \textbf{0} & \textbf{1} & \textbf{2} & \textbf{3} & \textbf{4} & \textbf{5+} & \textbf{Total} \\
        \hline
        \textbf{Number of clusters} & $73{,}751$ & $224{,}261$ & $2{,}765$ & $224$ & $51$ & $44$ & $301{,}096$ \\
        \textbf{Proportion of clusters} & $24.49\%$ & $74.48\%$ & $0.92\%$ & $0.07\%$ & $0.02\%$ & $0.01\%$ & $100.00\%$ \\
        \hline
    \end{tabularx}
    \caption{Distribution of the number of distinct SIRETs by cluster.}
    \label{tab:clusterSiretsDiff}
\end{table}

The remaining $3{,}084$ ($2 \%$) clusters contain two or more different SIRETs. This number must be interpreted while considering two additional points. First, a conflict does not necessarily concern the full SIRET: it is possible to have several SIRETs sharing the same SIREN (i.e. prefix) in a given cluster. This turns out to be the case for $1{,}408$ ($1 \%$) of the clusters. Second, a conflict is not necessarily due to the clustering step: it can be the result of an error occurring at the identification step, leading to some incorrect SIRETs. 
In order to assess specifically the clustering step, we focus only on the agents whose SIRETs are originally present in the TED, as we did in the \textit{Identification Method} section. We ignore these SIRETs in order to perform the clustering, and use them as ground truth. The results are presented in Table~\ref{tab:clusterFalsePositive}, and show that only $2.01 \%$ of the considered agents end up with a partially or completely incorrect SIRET.

\begin{table}[ht]
    \centering
    \begin{tabularx}{\textwidth}{X r r r r}
        \hline
        \textbf{Number of distinct known SIRETs} & \textbf{Full} & \textbf{Partial} & \textbf{Incorrect} & \textbf{Total}\\
        \hline
        \textbf{Number of clusters} & $23{,}762$ & $169$ & $393$ & $24{,}324$ \\
        \textbf{Proportion of clusters} & $97.99\%$ & $0.70\%$ & $1.31\%$ & $100.00\%$ \\
        \hline
    \end{tabularx}
    \caption{Distribution of the three possible outcomes of the clustering process when using the known SIRETs as ground truth.}
    \label{tab:clusterFalsePositive}
\end{table}

\subsubsection{False Negatives}
False negatives correspond to agents incorrectly placed in different clusters. In order to assess this type of error, we cannot use the same data as for the false positives, because the forms of these agents do not exhibit enough diversity. Instead, we constitute a more appropriate ground truth by sampling agents that appear several times under different forms. Our sample contains $5{,}020$ occurrences, that correspond to $377$ unique agents. Each one of the $377$ corresponding SIRETs appears between $1$ and $538$ times in the sample. For the sake of evaluation, we ignore these agents during the identification step, which is only applied to the rest of the data. We then perform the clustering step on the whole dataset, including the sample. Finally, we assess the false negatives produced during this last step by studying how the sampled agents are distributed over the clusters. For this purpose, we compute two measures proposed for the occasion.

\begin{figure}[ht]
    \centering
    \includegraphics[width=1\textwidth]{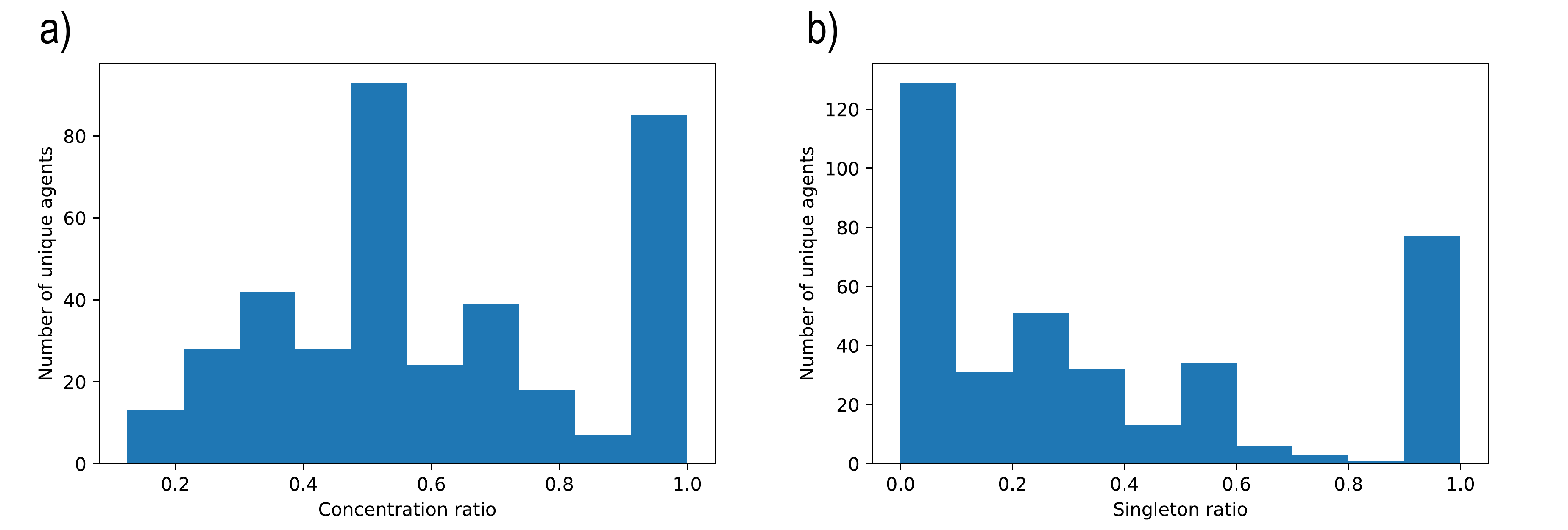}
    \caption{Distribution of the concentration ratio (a) and singleton ratio (b) over unique agents.}
    \label{fig:clusterHand}
\end{figure}

First, the \textit{Concentration Ratio} is defined for an agent of interest, and corresponds to the maximal proportion of occurrences of this agent present in a single cluster. A value of one indicates that all the occurrences belong to the same cluster, whereas a low ratio shows that the occurrences are scattered over a number of clusters. Figure~\ref{fig:clusterHand}a shows the distribution of the concentration ratio over unique agents. The mean concentration ratio is $0.6$, which means that more than half of an agent's occurrences are placed in the same cluster, in average. Our method perfectly clusters $243$ agent occurrences ($5 \%$), representing $83$ unique agents ($22 \%$). These are clusters with $2$ or $3$ occurrences: it is apparently hard to gather many occurrences of the same agent in a single cluster. The others agents are less concentrated, with a majority of clusters containing half of their occurrences. 

Second, the \textit{Singleton Ratio} corresponds to the proportion of the agent's number of occurrences forming singleton clusters. A value of one indicates that all the agent's occurrences are distributed in their own cluster, whereas a low ratio shows that the occurrences belong to larger clusters. Figure~\ref{fig:clusterHand}b shows the distribution of the singleton ratio over unique agents. On the one hand, there are $77$ unique agents ($20 \%$) with a ratio of $1$. This means that all related occurrences are located in separate clusters. This is the case for occurrences that are very different, for example that do not share the same name. This result confirms the relevance of our identification process: some occurrences can only be gathered by finding the correct SIRET number. On the other hand, the ratio of the rest of the agents is much lower. The mean singleton ratio for them (i.e. excluding ratios equal to 1) is $0.18$. This means that our method splits the set of occurrences linked to a SIRET into only a very limited number of subgroups.


\subsection{Overall Assessment}
Overall, our method allows significantly improving the quality of the data. Before its application, the SIRET is filled for only $16.4 \%$ and $2.9 \%$ of the unique buyers and winners, respectively. Afterwards, the completion rates for the same fields are $97.5 \%$ and $86.2 \%$. Assuming that the results obtained on the different samples for agent identification and merging are representative of the whole dataset, we can estimate the performance for agent occurrences, at each step of the process.

\begin{table}[ht]
    \centering
    \begin{tabularx}{\textwidth}{X r r r r r r}
        \hline
        \textbf{Step} & \multicolumn{2}{r}{\textbf{Correct}} & \multicolumn{2}{r}{\textbf{Incorrect}} & \multicolumn{2}{r}{\textbf{Missing}} \\
        \hline
        After Separation      &     $247{,}350$ &  $8.59\%$ &         $0$ &  $0.00\%$ & $~~~~~~2{,}631{,}333$ & $91.41\%$ \\
        After Normalization   &     $649{,}170$ & $22.55\%$ &         $0$ &  $0.00\%$ & $2{,}229{,}513$ & $77.45\%$ \\
        After Identification  & $2{,}311{,}418$ & $80.29\%$ & $363{,}497$ & $12.63\%$ &     $203{,}768$ &  $7.08\%$ \\
        After Clustering      & $~~~~~~2{,}313{,}622$ & $80.37\%$ & $~~~~~~363{,}533$ & $12.63\%$ &     $201{,}528$ &  $7.00\%$ \\
        \hline
    \end{tabularx}
    \caption{Number of agent occurrences counted originally, and after each step of the proposed process.}
    \label{tab:AgentStats}
\end{table}

Table~\ref{tab:AgentStats} shows the evolution of the number of agent occurrences, depending on their identification status (correct, incorrect, none), after the main steps of our pipeline. As a baseline, we originally have $2{,}761{,}930$ agent occurrences before any processing, as identified through exact matching. The first row (\textit{After Separation}) shows the situation after having separated multiple agent occurrences (cf. Section \textit{Database Initialization}). At this stage, all the present identifiers are correct, but this concerns a very small part of the now $2{,}878{,}683$ agent occurrences. The second row (\textit{After Normalization}) shows the number of agents after having normalized their names. This step matches occurrences that correspond to the same entity, some of which have a SIRET, thereby allowing to identify an additional $14 \%$ of the total occurrences. The third row (\textit{After Identification}) focuses on the retrieval of missing identifiers by leveraging the national database. This step clearly has the strongest effect, with an additional $70 \%$ of identified agent occurrences. However, this does not go without the introduction of some incorrect SIRETs among them. Finally, the last row (\textit{After Merging}) shows the results after the merging of certain agents. The effect is clearly smaller, but it still allows identifying more than $2{,}000$ additional occurrences.

A manual analysis of the agents that our process cannot identify, or identifies incorrectly, reveals that they correspond to situations where agent information is so incomplete and/or incorrect that even handling them manually proves to be a very difficult, or even impossible, task. For instance, $71 \%$ of the agents remaining unidentified at the end of the process have no location information whatsoever.

\section{Usage Notes}
As explained before, FOPPA was developed in the framework of the DeCoMaP project with two goals in mind: perform a descriptive analysis of public procurement data in order to better understand its functioning and identify problems or anomalies; and train and test an automatic method able to detect frauds. Obviously, FOPPA can therefore be used for these tasks. It is worth stressing that the fraud detection method proposed in the context of DeCoMaP relies on a graph-based model of public procurement, in which vertices represent buyer and winners, and edges the contracts involving them~\cite{Potin2023,Potin2023b}. To build such graphs, it is necessary to be able to identify economic agents in a unique way. This is the reason why we spent a lot of time and efforts fixing the missing SIRETs problem in the raw data. But of course, FOPPA can be used to train and/or test methods that are not based on graphs.

More traditional methods to detect fraud in public procurement involve the identification of red flags based on some basic statistics. However, these also require an accurate identification of both the winning companies and the buyers. For example, Stefanov \textit{et al}. build an Herfindhal-Hirsh index of concentration for buyers based on relative share of contractors~\cite{Stefanov2015bulgarian}. Similarly, Fazekas \& Kocsis construct a corruption red flag based on evaluation criteria~\cite{Fazekas2020uncovering} (sum of weights for criteria which are not related to prices and quantitative requirements). As the authors notice, ``in TED, information on award criteria was available in an unstructured text variable along with the weight of each criterion'' and  require, for statistical use, a reprocessing like the one proposed in FOPPA. It is also possible to build red flags by combining standard indicators extracted from public databases like FOPPA, with others based on data that are not publicly available, as Decarolis \& Giorgiantonio do with information describing police and judiciary operative practices~\cite{Decarolis2022}. Beside fraud detection, Klingler stresses the need for regulatory impact assessments to evaluate the economic consequences of public procurement regulation~\cite{Klingler2020measuring}. But such impact studies cannot be performed without a database that can be relied upon.

It should be stressed that, as most of our efforts have mainly focused on the identification of buyers and suppliers, some limitations of the original data are still present in FOPPA. First, due to the incompleteness of the raw TED data, many notices are missing from FOPPA: $39 \%$ of contract notices and $19 \%$ of award notices miss their counterparts. These numbers should be downplayed, though, as they include two cases that are unfortunately indistinguishable in the TED: not only errors (notices that should be in the dataset), but also licit cases (optional notices). Also, most of the contract information appears in \textit{both} types of notices, the award notices being the most informative overall. As mentioned in the \textit{Main Issues} section, this is a well-known problem of the TED~\cite{TED2020b, Csaki2018}, whose solving would require leveraging a more complete alternative data source, which as far as we know, is not publicly available. 
The second limitation concerns the incompleteness of certain important fields describing the lots. The contract duration and advertisement period are missing in $39 \%$ and $25 \%$ of the lots, respectively. This is mainly due to the missing contract notice issue, as both of these fields do not appear in the award notices. The number of tenders, which is used to define certain red flags, is missing for $31 \%$ of the lots. Finally, the contract price, a crucial piece of information, is missing in $31 \%$ of the lots. Moreover, the prices indicated in the rest of the lots are known to be unreliable, as mentioned in several reports on TED data quality~\cite{Ackermann2019, EU2020}. This is also a problem with the raw data, that would require an alternative data source to be solved.

A preexisting database called Opentender~\cite{Opentender2018} and developed in the framework of the Digiwhist~\cite{Digiwhist2018} project, is comparable to FOPPA in the sense that it mainly focuses on public procurement notices published on the TED. However, Opentender is much more ambitious in terms of coverage, as it contains data describing 25 European countries, and includes individuals in addition to institutions and companies. For this reason, it required combining a number of very heterogeneous data sources besides the TED. Constituting such a database is a very difficult task, and in practice it is impossible to tailor the proper processing required by each considered source. Consequently, the quality of the French data in Opentender does not reach the level obtained in FOPPA. When focusing on the notices present in both databases, we identified four issues in Opentender compared to FOPPA. We summarize them here, but the interested reader will find a detailed description in our technical report~\cite{Potin2022}. First, $14\%$ of the lots present in FOPPA (and the TED) are missing in Opentender. Second, Opentender contains roughly \textit{twice} as many unique agents as FOPPA, because many similar (but not identical) occurrences are considered as distinct, when they actually represent the same entity. It should be noted that the identification of unique agents was not as critical for the Digiwhist project as it is for DeCoMaP. Third, the contracts not awarded to any supplier are processed like the successful ones, resulting in more than $4\%$ of spurious agents among Opentender winners (often called ``Infructeux'', French for ``Unsuccessful''). Fourth, the award criteria are missing in $36\%$ of the considered notices, when they are present in the TED and FOPPA. This highlights the importance of the efforts we put in adapting our processing to the specific case of the French notices. It is important to stress that our work is meant to be used by others. The FOPPA data could be integrated to Opentender or any other similar database. The data processing method that we propose is generic enough to be adapted to other similar national data sources in a relatively straightforward way, allowing to deal with the remaining member states.

An interesting characteristic of FOPPA is that, thanks to our efforts, a large part of the economic agents it contains are identified by their SIRET number. This identifier is ubiquitous in French data, both in open and commercial sources, and therefore allows connecting FOPPA to a number of existing databases, thereby opening new perspectives. First, one can retrieve a number of additional details concerning the agents in the SIRENE database~\cite{Sirene2021} itself (e.g. activity domain, number of employees) or in its extensions such as the geolocated version~\cite{Sirene2022} (GPS coordinates). Second, certain types of agents have their own databases containing type-specific characteristics, e.g. high schools~\cite{Idf2021}, parkings~\cite{Pan2020}, cinemas~\cite{Osm2020}, operators committed to organic farming~\cite{AgenceBio2023}. Third, some available databases reflect the functioning of the economic agents, e.g. financial aggregates of the municipal accounts~\cite{Ofgl2021}, union election results~\cite{MinTrav2019}, collective agreements declared by companies~\cite{MinTrav2023}, index of professional equality~\cite{MinTrav2023a}. Fourth and finally, certain databases allows connecting the \textit{legal} entities from FOPPA to some \textit{natural} persons such as beneficial owners~\cite{Inpi2022} or elected representatives~\cite{MinInt2018}.

\section{Code Availability}
Our Python source code is publicly available online\footnote{\href{https://github.com/CompNet/FoppaInit/tree/v1.0.2}{\texttt{https://github.com/CompNet/FoppaInit}}} as a GitHub repository~\cite{Potin2022b}. It is designed to be applied to the raw TED tables~\cite{Ted2010, Ted2010a, Ted2011, Ted2011a, Ted2012, Ted2012a, Ted2013, Ted2013a, Ted2014, Ted2014a, Ted2015, Ted2015a, Ted2016, Ted2016a, Ted2017, Ted2017a, Ted2018, Ted2018a, Ted2019, Ted2019a, Ted2020, Ted2020a}, and leverages the Hexaposte~\cite{LaPoste2022} and SIRENE~\cite{Sirene2021} databases mentioned in the \textit{Methods} section. It performs the integrality of the processing described in this section, and produces the FOPPA database. When performed in parallel on 10 NVIDIA GeForce RTX 2080 Ti GPUs, this processing requires approximately 6 days.

\section{Acknowledgements}
This work was supported by the \textit{Agence Nationale de la Recherche} (ANR, France) under grant number ANR-19-CE38-0004 for the \textit{DeCoMaP} project. This project was originally initiated thanks to the support of the \textit{Agorantic} research federation (FR 3621).

\section{Author Contributions Statement}
L.P. retrieved the raw datasets, wrote the source code and constituted the database. All authors participated in designing the data processing, writing, reviewing, and revising the manuscript.

\section{Competing Interests}
The authors declare no competing interests.

\phantomsection\addcontentsline{toc}{section}{References}
\printbibliography


\end{document}